\documentclass{optica-article}

\journal{opticajournal} 

\articletype{Research Article}

\usepackage{siunitx}
\usepackage[position=top]{subfig}
\usepackage{lineno}
\usepackage{booktabs}

\usepackage{physics}

\begin{document}

\title{Programmable time-frequency mode-sorting of single photons with a multi-output quantum pulse gate}

\author{L. Serino\authormark{1,2,*}, C. Eigner\authormark{2}, B. Brecht\authormark{1,2}, C. Silberhorn\authormark{1,2}}
\address{\authormark{1}Paderborn University, Integrated Quantum Optics, Warburgerstr. 100, Paderborn, Germany}
\address{\authormark{2}Paderborn University, Institute for Photonic Quantum Systems, Warburgerstr. 100, Paderborn, Germany}
\email{\authormark{*}laura.serino@upb.de}

\begin{abstract*}
We demonstrate a high-dimensional mode-sorter for single photons based on a multi-output quantum pulse gate, which we can program to switch between different temporal-mode encodings including pulse modes, frequency bins, time bins, and their superpositions. This device can facilitate practical realizations of quantum information applications such as high-dimensional quantum key distribution, and thus enables secure communication with enhanced information capacity. We characterize the mode-sorter through a detector tomography in 3 and 5 dimensions and find a fidelity up to $0.958\pm 0.030$ at the single-photon level.
\end{abstract*}

\section{Introduction}
Quantum information science based on high-dimensional encodings has the potential to revolutionize computing, communication, and cryptography by harnessing the unique properties of quantum systems \cite{kimble08, obrien09}. 
High-dimensional encodings allow for increased information capacity of quantum information carriers and noise robustness, thereby improving the security and efficiency of quantum communication protocols such as high-dimensional quantum key distribution (HD-QKD) \cite{sheridan10, ecker19}. 
Photons are a natural choice of information carrier due to their inherent quantum properties and high-dimensional degrees of freedom in both the spatial and time-frequency domain. 
Although arguably less explored than their spatial counterparts, time-frequency encodings offer significant advantages in quantum information applications: since orthogonal time-frequency modes can share the same spatial distribution, they are resilient in transmission and compatible with existing spatially single-mode telecom fiber infrastructure.

Temporal modes (TMs) provide appealing time-frequency encodings in the form of field-orthogonal wave-packet modes that can be expressed as coherent superpositions of monochromatic modes \cite{brecht15}.
These encodings can describe discrete Hilbert spaces with finite high dimensionality, which find applications in many quantum information fields, from quantum cryptography \cite{sheridan10} to quantum computing \cite{lanyon09, presutti24} and quantum networks \cite{guha11, rosati16, folge24}.
TM alphabets used in these applications include frequency bins, time bins, and pulse modes, each offering distinct advantages and posing unique challenges \cite{brecht15, raymer20, lu23b}. 
Frequency bins and time bins encode information in modes that are, for all practical purposes, intensity-orthogonal in frequency or time, respectively. In contrast, the intensity profiles of pulse modes overlap in time and frequency, and the field-orthogonality condition is determined by their complex spectral profile. Hermite-Gauss (HG) modes are a common example of this type of encoding alphabet.

Despite their different labels, frequency bins, time bins, and pulse modes all belong to the same family: they can be described as complex functions of time or frequency, connected through a Fourier transform and, therefore, they all represent particular realizations of TMs.
We note that, since TMs are defined as wave-packet modes, they differ conceptually from the more conventional continuous variable approach which uses time and frequency as conjugate bases containing an infinite number of delta-like bins \cite{qi06, mower13, nunn13, widomski23}. Instead, TMs include discrete finite bins and pulse modes that can be encoded in light pulses. Conjugate bases in this context are constructed as superpositions of states from the fundamental basis with appropriate phases.

Establishing a complete TM-based quantum information framework requires the generation, manipulation, and detection of high-dimensional quantum TM states. 
Engineered parametric down-conversion is a well-known and established tool for generating these states \cite{morrison22, chiriano23, serino24}. 
However, simultaneously manipulating or detecting multiple single-photon TMs is challenging because both pulse modes and superpositions of bins overlap in time and frequency. 
This task is further complicated by the requirements of quantum communication protocols like HD-QKD, which demand a high-dimensional mode sorter capable of operating with single photons in real time, that is, on a shot-by-shot basis.

To date, dedicated devices have been developed and tailored for each application. The so-called quantum frequency processor \cite{lu18a, lu23a}, for instance, demonstrated mode-sorting of three-dimensional frequency bins and their superpositions using a combination of phase modulators and pulse shapers. Similarly, interferometric setups based on beam splitters \cite{vagniluca20} or group-velocity dispersion \cite{widomski24} have been used to decode time bins in up to four dimensions and two conjugate bases. 
These solutions, however, rely on complex setups which hinder scalability to a larger number of dimensions or measurement bases and lack reconfigurability. 
The multi-output quantum pulse gate (mQPG) \cite{serino23} has recently shown simultaneous projections of a single-photon-level input state onto five arbitrary superpositions of pulse modes using a single nonlinear process, which can be scaled to higher dimensions without increasing the number of components. 
However, no analogous demonstration yet exists for frequency bins and time bins.
The need for dedicated infrastructure for each encoding complicates interfacing between devices that work with different alphabets, even though they operate on the same degree of freedom.

In this work, we demonstrate programmable high-dimensional mode-sorting of different TM-based encoding alphabets at the single-photon-level, achieved through an mQPG. This programmability is enabled through spectral pulse shaping, leveraging the fact that all TM alphabets can be described as complex functions of frequency to change the measurement basis or encoding alphabet without requiring hardware modifications. For each encoding alphabet --- pulsed modes, frequency bins and time bins --- we demonstrate high-fidelity projections onto multiple superposition bases, namely all possible mutually unbiased bases (MUBs) in 3 and 5 dimensions. 
The versatility of the mQPG to operate in different high-dimensional encoding alphabets and superposition bases is crucial for interfacing with diverse sources and integrating the unique benefits offered by each encoding alphabet, paving the way for practical applications in quantum information science.

\begin{figure}
    \centering
    \includegraphics[width=.9\linewidth]{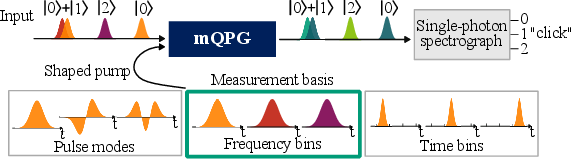}
    \label{fig:concept}
    \caption{Working principle of the mQPG-based time-frequency mode sorter: the mQPG process, driven by a pump pulse spectrally shaped to implement the selected measurement basis, up-converts the input photons to a different output frequency based on their temporal mode; the output frequencies, which encode the information in the signal, can be detected with a single-photon spectrograph. We can choose frequency bins, pulse modes, and time bins as encoding alphabet.}
\end{figure}

\section{The multi-output quantum pulse gate}
\label{sec:theory}
The mQPG serves as a high-dimensional mode-sorter for TM states \cite{serino23}. Its working principle is based on a dispersion-engineered sum-frequency-generation (SFG) process where an input pulse, driven by a strong pump field, is converted into a particular output pulse based on its TM. 
The process takes place in a periodically poled nonlinear waveguide, typically titanium in-diffused lithium niobate \cite{brecht14}.
The periodic poling structure of the mQPG waveguide, consisting of alternating poled and unpoled regions (``super-poling''), enables parallel SFG processes centered at distinct output frequencies which define the output channels of the device.
The SFG process is engineered to achieve group-velocity matching between the input and pump pulses, which is essential for mode-selective operation in each channel \cite{eckstein11}. 

\begin{figure}
    \centering
    \includegraphics[]{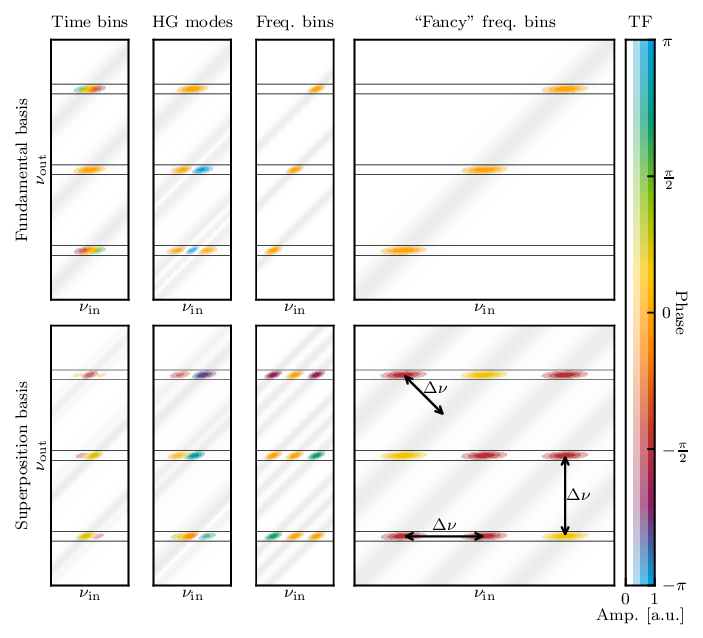}
    \caption{Frequency-space representation of the transfer functions of the three-dimensional fundamental basis (top) and a superposition basis (bottom) for the three TM alphabets tested in this work: time bins, HG modes, and frequency bins (including the ``fancy'' frequency bin mode-sorting approach explained in the main text). The horizontal lines indicate the phase-matching function $\Phi(\nu_\mathrm{out})$, comprising one phase-matching peak for each output channel of the mQPG. The diagonal gray area shows the amplitude of the pump function $\alpha(\nu_\mathrm{out} - \nu_\mathrm{in})$. The colored regions indicate the transfer function $G(\nu_\mathrm{in}, \nu_\mathrm{out})$, which shows where the SFG process can take place.}
    \label{fig:tf}
\end{figure}

In frequency space, the mQPG operation is described by a transfer function $G(\nu_\mathrm{in}, \nu_\mathrm{out})$ that relates the input frequencies $\nu_\mathrm{in}$ to the output frequencies $\nu_\mathrm{out}$ (Fig.~\ref{fig:tf}). 
The transfer function is the product of the pump function $\alpha$, which describes energy conservation, and the phase-matching function $\Phi$, which is determined by momentum conservation: $G(\nu_\mathrm{in}, \nu_\mathrm{out}) = \alpha(\nu_\mathrm{out} - \nu_\mathrm{in}) \cdot \Phi(\nu_\mathrm{out})$.
The number and spacing of phase-matching peaks depend on the super-poling parameters selected during the fabrication of the mQPG waveguide \cite{serino23}. 
The pump function is determined by the complex spectral amplitude of the pump pulse, which can be tailored through spectral shaping. By dividing the pump spectrum into distinct spectral regions, spaced to match the phase-matching peaks, each region can be aligned with a specific phase-matching peak in correspondence of the same input frequencies (as shown in the first three columns of Fig.~\ref{fig:tf}). Each spectral region can then be shaped to implement a different TM in the respective output channel, effectively creating a multi-output transfer function.

If the spectral features of the pump are larger than the bandwidth of each phase-matching peak, the mQPG performs ideal single-mode projections in each channel, meaning that the up-conversion probability in each output channel is proportional to the complex spectral overlap between the corresponding pump mode and the input state \cite{donohue18}.
In this optimal regime, each mQPG channel can be described by a von-Neumann projection $\pi^\gamma = \ket{\gamma}\bra{\gamma}$ onto the assigned TM $\gamma$. In this case, the probability of SFG conversion of a pure input state $\rho^\xi = \ket{\xi}\bra{\xi}$ is given by $p^{\gamma\xi} = \Tr{\rho^\xi\pi^\gamma}$ \cite{brecht14}. 
The complete mQPG process, described by the set of projectors $\{\pi^\gamma\}$ comprising all channels, effectively sorts the input modes into output frequencies which can be read out using a spectrograph.
The modes $\gamma$ for the projections can be chosen programmatically via pump shaping, which facilitates measurements in different superposition bases without any hardware modifications. Changing the encoding alphabet is equally straightforward, requiring only a change in the shaped pulse from frequency bins to pulse modes or time bins or any of their superpositions.

In a practical experimental setting, however, the mQPG has a non-unity conversion efficiency, which leads to an inconclusive result when the input photons are not up-converted by the device. 
In this case, the mQPG operation is described by the positive-operator-valued measure (POVM) $\{\pi^\gamma\}$, comprising the POVM elements $\pi^\gamma$ that describe the operation in each channel and an additional POVM element describing the inconclusive result. For simplicity, in the following discussion we will neglect the inconclusive result and consider only the POVM elements associated with successful click detection.
Experimental constraints can also limit the achievable ratio between the spectral bandwidth of the pump and that of the phase-matching bandwidth, introducing spurious multi-mode processes in each channel that lead to cross-talk in the mode-sorting process \cite{donohue18}. In this case, the POVM elements require the more general description $\pi^\gamma = \sum_{ij}{m^\gamma_{ij}\ket{i}\bra{j}}$, with $i$ and $j$ elements of the chosen fundamental TM basis \cite{ansari17}. 
In order to eliminate the induced cross-talk, one must artificially reduce the phase-matching bandwidth through narrowband frequency filtering. 
However, the resolution of currently available single-photon spectrographs in the mQPG output frequency range (around \SI{550}{\nano\meter}) is insufficient for adequate frequency filtering. This technical limitation significantly constrained previous demonstrations, decreasing the measurement fidelity of the mQPG in five dimensions from the intrinsic value of 92\% to 65\%\footnote{These values have been adapted to the definition of fidelity used in this work, which differs from the one in \cite{serino23} by a square root.} \cite{serino23}.

To overcome this limitation, we developed an alternative technique for frequency-bin mode-sorting, labeled ``fancy'' frequency bin (FFB) approach, that leverages the periodic phase-matching structure of the mQPG to improve its performance. Indeed, it is immediate to notice that a single pump bin intercepts multiple phase-matching peaks, separated by $\Delta\nu$, in correspondence of input frequencies also separated by $\Delta\nu$ (last column in Fig.~\ref{fig:tf}). This transfer function, generated using only one pump bin, can completely mode-sort discrete frequency bins with separation $\Delta\nu$ in input, which form the fundamental basis of a $d$-dimensional Hilbert space. 

This scheme can be intuitively extended to also mode-sort MUBs containing superpositions of frequency bins by generating pump bins with spacing equal to $\Delta\nu$. In contrast to the standard approach, where each pump bin addresses only one mQPG channel, in the FFB method each pump bin intercepts multiple phase-matching peaks at corresponding input frequencies with an offset of $\Delta\nu$. A total of $2d-1$ pump bins is thus sufficient to generate a complete $d\cross d$ transfer function, where the mode in each mQPG channel is determined by the relative phase of the pump bins. This approach enables sorting $d$ distinct input modes (frequency bins or their superpositions) starting from only $2d-1$ pump frequency bins, instead of the $d^2$ bins required by the standard approach. Hence, the FFB method scales linearly with the system dimensionality rather quadratically, effectively reducing the technical demands on the experimental setup. Namely, it allows one to use fewer, broader pump bins within a fixed spectral region, reducing the need for precise spectral shaping while simultaneously increasing the ratio between the spectral features of the pump and the phase-matching width. This generally results in lower cross-talk \cite{donohue18}, greatly improving the measurement quality for a fixed available pump bandwidth and reducing reliance on extensive output filtering.

While the FFB approach introduces inter-bin correlations that restrict the full programmability of each channel, it still provides sufficient flexibility for various applications. For instance, in prime dimensions $d$, this technique allows for measuring in $d$ MUBs \cite{combescure09}, only one basis short of the complete set of $d+1$ MUBs. Although this limitation precludes applications such as input state tomography, which require measurements in all MUBs, the FFB method offers significant advantages in HD-QKD, which only requires at least two MUBs and can strongly benefit from additional measurement bases.

\section{Experiment}

\begin{figure}
    \centering
    \includegraphics[width=0.9\linewidth]{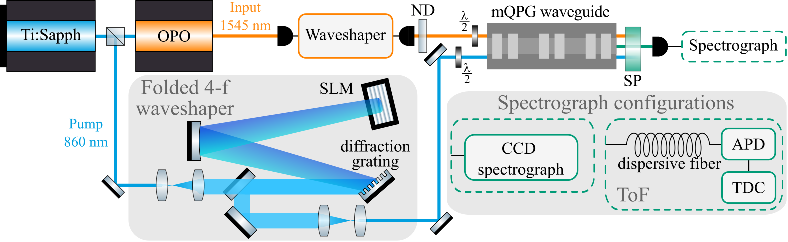}
    \caption{Schematic of the experimental setup. A system comprising a mode-locked Ti:sapphire laser and an optical parametric oscillator (OPO) generates the initial input and pump pulses centered at \SI{1545}{\nano\meter} and \SI{860}{\nano\meter}, respectively. The input pulse is attenuated to $\langle \hat{n} \rangle < 0.1/$pulse using a neutral density (ND) filter and shaped by a commercial waveshaper to generate the input mode. The pump modes are shaped using a folded 4-f waveshaper \cite{monmayrant10} comprising a diffraction grating, a cylindrical mirror and a spatial light modulator (SLM). After waveshaping, input and pump pulses are both coupled into the mQPG waveguide. The up-converted output photons are isolated by a short-pass (SP) filter and detected with a CCD spectrograph or a time-of-flight (ToF) spectrograph \cite{avenhaus09} formed by a dispersive fiber, an avalanche photodiode (APD) and a time-to-digital converter (TDC).}
    \label{fig:setup}
\end{figure}

The schematic of the experimental setup is illustrated in Fig.~\ref{fig:setup}. A mode-locked Ti:sapphire oscillator with a repetition rate of \SI{80}{\mega\hertz} generates pulses centered at \SI{860}{\nano\meter} with a spectral full-width-half-maximum of \SI{3}{\tera\hertz}, corresponding to a duration of \SI{150}{\femto\second}. The pump pulse is generated by shaping the spectral amplitude and phase of the laser pulse through a in-house-built 4-f-line waveshaper based on a spatial light modulator.
A portion of the original Ti:sapphire pulse is directed to pump an optical parametric oscillator that generates the input pulse of the experiment centered at \SI{1545}{\nano\meter}, which is shaped by a commercial waveshaper and attenuated to below 0.1 photons per pulse to generate the single-photon-level input states, representing the typical information carrier of a quantum communication scheme such as HD-QKD.
For this demonstration, we consider three TM alphabets: time bins, HG modes, and frequency bins. For the latter, we also test the alternative FFB mode-sorting technique explained in the previous section by appropriately shaping the pump pulse. The spectral parameters used for each encoding alphabet are reported in Table \ref{tab:param}, and examples of the resulting spectra are shown in the Supplementary Material. 

The pump and input pulses are coupled into the mQPG, realized as a periodically super-poled titanium-indiffused lithium niobate waveguide \cite{serino23}. 
The poling period of \SI{4.32}{\micro\meter} allows for a type-II SFG process in which a horizontally-polarized \SI{1545}{\nano\meter} input photon is up-converted to a horizontally-polarized output photon at \SI{552.5}{\nano\meter} through a vertically-polarized \SI{860}{\nano\meter} pump field, according to the mQPG mode-sorting process.
We use a five-output mQPG waveguide of which we use 3 or 5 output channels, depending on the dimensionality of the chosen alphabet. The output channels are centered at frequencies separated by \SI{0.5}{\tera\hertz} in the case of FFB in 5 dimensions and \SI{0.63}{\tera\hertz} in all other cases.
The output field is isolated from the leftover pump and input fields via frequency filtering and detected with an in-house-built time-of-flight (ToF) spectrograph consisting of a dispersive fiber, an avalanche photodiode, and a time-to-digital converter \cite{avenhaus09}. 
The ToF spectrograph facilitates shot-to-shot measurements (required, e.g., by HD-QKD protocols) and is cost-effective as it utilizes only one single-photon detector. Nevertheless, technical constraints in the current implementation limit its effective resolution to \SI{300}{\giga\hertz}, a much greater value than the bandwidth of each phase-matching peak (approximately \SI{30}{\giga\hertz}) which negatively affects the overall quality of the measurements.
Therefore, to showcase the quality of the mQPG operation when paired to optimal frequency-resolving measurements in output, we repeat the measurements replacing the ToF spectrograph with a CCD spectrograph (Andor Shamrock 500i) with a resolution of \SI{30}{\giga\hertz}. We note that this device is generally unsuitable for shot-to-shot measurements as it requires a read-out time of the order of \si{\milli\second}, incompatible with the \SI{80}{\mega\hertz} repetition rate of the laser.

For each tested encoding alphabet and dimensionality, we quantify the measurement quality of the mQPG through a quantum detector tomography \cite{lundeen09, ansari17, serino23}. 
This technique aims at reconstructing the POVMs $\{\pi^\gamma\}$ by probing the device with an informationally complete set of input states --- in this case, all eigenstates of all possible MUBs in the chosen Hilbert space --- and counting the output photons in each channel.

To perform the quantum measurement tomography, we first choose the measurement basis that we want to probe, which defines the POVM $\{\pi^\gamma\}$ comprising the POVM elements of each mQPG channel, and we set up the mQPG mode-sorting process by shaping the pump spectrum accordingly. We probe the mQPG with input states from the complete set $\{\rho^\xi\}$ of the $d(d+1)$ elements of all $d+1$ MUBs of the considered $d$-dimensional Hilbert space. 

For each measurement we run 10-20 acquisitions, each with an integration time between \SI{0.5}{\second} and \SI{1}{\second}, chosen based on the measured count rate. The input power is always adjusted to have $\langle \hat{n} \rangle < 0.1/$pulse, whereas the pump power is maximized within the capabilities of the experimental setup, reaching \SIrange{5}{10}{\milli\watt}. The maximum available pump power, which determines the conversion efficiency and therefore the number of output counts in each measurement, depends on the chosen encoding alphabet and dimensionality, as the waveshaping system needs to ``carve'' the appropriate spectrum from the original laser pulse.

We normalize the measured counts to obtain the experimental output probabilities for each channel $p^{\gamma\xi}$. From these values, we reconstruct the POVM elements $\pi^\gamma$ through a weighted least-squares fit by minimizing the quantity
$\sum_\xi{{\left|p^{\gamma\xi} - \Tr{\rho^\xi\pi^\gamma}\right|^2}/{p^{\gamma\xi}}}$,
where we constrain $\pi^\gamma$ to be Hermitian and positive semidefinite.
We then compare the reconstructed POVM elements to the theoretical projectors $\ket{\gamma}\bra{\gamma}$ by calculating the fidelity $\mathcal{F} = \bra{\gamma}\pi^\gamma\ket{\gamma} / \mathrm{Tr}(\pi^\gamma)$.

\section{Results and discussion}
\begin{figure}
    \centering
    \includegraphics[]{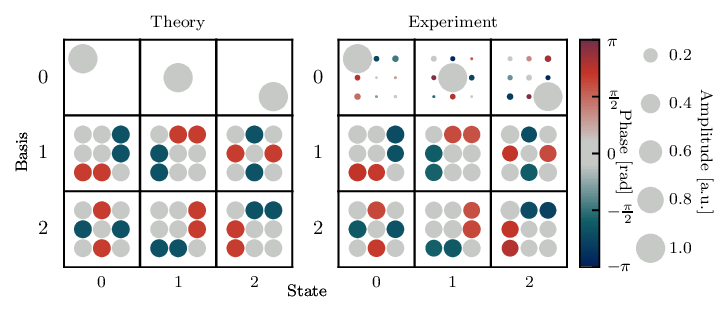}
    \caption{Theoretical (left) and experimentally characterized (right) POVMs for three-dimensional operation in the frequency-bin alphabet (FFB scheme) at the single-photon level ($\langle \hat{n} \rangle < 0.1/$pulse).}
    \label{fig:povms}
\end{figure}

Through the detector tomography, we reconstruct the POVMs that describe the mQPG operation when projecting an input state into the selected MUB. For each encoding alphabet, we test all $d+1$ MUBs in $d=3$ and $d=5$, with the exception of the FFB scheme, which is limited to $d$ MUBs. 
Figure \ref{fig:povms} shows the reconstructed POVMs compared to the theoretical ones for the FFB scheme in $d=3$, corresponding to a measurement at the single-photon level acquired with the ToF spectrograph.
We highlight that each row corresponds to 3 POVM elements that are measured simultaneously and, therefore, form the POVM that describes the mode-sorting operation of the mQPG for the respective basis.

\begin{figure}
    \centering
    \includegraphics[]{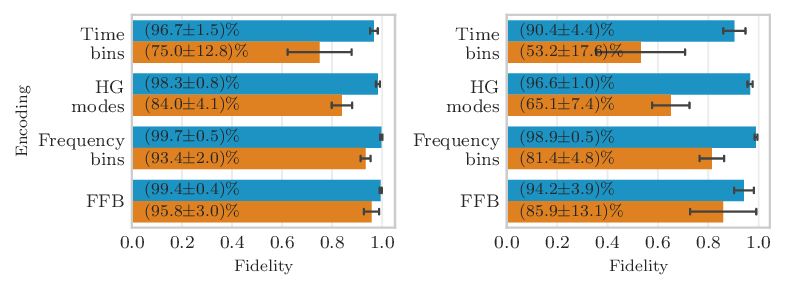}
    \caption{Average fidelity of the measured POVMs to the ideal operators for each encoding alphabet: time bins, HG modes, and frequency bins both with the standard and FFB approach detailed at the end of section \ref{sec:theory}. The blue bars represent the results of the measurements performed with the CCD spectrograph in output, corresponding to ideal filtering conditions, whereas the orange bars show the results of the ToF detection at the single-photon level ($\left< \hat{n} \right> \leq 0.1$/pulse). The error bars indicate the standard deviation of the fidelity for different POVMs.}
    \label{fig:fidelity}
\end{figure}

Figure \ref{fig:fidelity} shows the average fidelity of the reconstructed POVMs to the ideal operators for all tested TM alphabets in 3 and 5 dimensions. 
The fidelity measured with the CCD spectrograph (blue bars) is always above 90\% and, for frequency bins, reaches (99.7$\pm$0.5)\% in $d=3$ and (98.9$\pm$0.5)\% in $d=5$, confirming the high intrinsic quality of the mQPG projections. 
This value is generally higher than the fidelity measured with the ToF spectrograph (orange bars), showing the benefits of optimal filtering in the output of the mQPG to enhance the quality of the mode-sorting process.
This effect is more noticeable in functions with narrow spectral features, such as superpositions of time bins or HG modes (shown in the Supplementary Material), especially at higher orders, which benefit more from spectral filtering to reduce cross-talk. Additionally, HG modes and frequency bins in $d=5$ need to be rescaled to fit within the same spectral bandwidth, as higher-order modes become spectrally broader. 

Frequency bins, however, maintain the same spectral features when building MUBs, and only change the relative phase between the different bins.
This results in generally higher fidelity in all bases, visible in the high mean value and narrow spread in Fig.~\ref{fig:fidelity}.
The FFB approach increases the spectral bandwidth available for each mode, reducing measurement imperfections and, consequently, enhancing resilience to the low-resolution ToF spectrograph in output.

Nevertheless, the spectrograph must resolve the output frequencies corresponding to the different channels of the mQPG, separated by \SI{0.5}{\tera\hertz} or \SI{0.63}{\tera\hertz} in this work. 
If the spectrograph resolution is not significantly larger than this value, the readout system introduces cross-talk into the measurement. In fact, this additional cross-talk is the main source of error for the highest-fidelity mode-sorting in $d=5$ presented in this work, obtained with the FFB approach. 
Moreover, the larger spectral bandwidth achieved with this technique leads to narrower time durations, which necessitate precise synchronization of the pump and input in time and compensation for delay drifts caused, e.g., by temperature fluctuations.
Correcting the delay of the pulses and improving the spectrographic detection will increase stability and reduce errors, facilitating the scalability of the mQPG mode-sorter to even higher dimensions. Optimizing the spectral parameters for each encoding and dimensionality will further enhance the measurement fidelity, enabling high-quality programmable mode-sorting in a large number of high-dimensional bases.

\begin{table}
    \centering
    \begin{tabular}{l|cc|cc}
        \toprule
        & \multicolumn{2}{c|}{$\mathbf{d=3}$} & \multicolumn{2}{c}{$\mathbf{d=5}$} \\
        \textbf{Encoding} & FWHM & Separation & FWHM & Separation \\
        \midrule
        Time bins & \SI{1.5}{\pico\second} & \SI{3.5}{\pico\second} & \SI{1.5}{\pico\second} & \SI{5.0}{\pico\second}\\
        HG modes & \SI{210}{\giga\hertz} & - & \SI{140}{\giga\hertz} & - \\
        Frequency bins & \SI{100}{\giga\hertz} & \SI{200}{\giga\hertz} & \SI{50}{\giga\hertz} & \SI{100}{\giga\hertz}\\
        FFB & \SI{300}{\giga\hertz} & \SI{630}{\giga\hertz} & \SI{150}{\giga\hertz} & \SI{500}{\giga\hertz} \\
        \bottomrule
    \end{tabular}
    \caption{Spectral parameters used for each encoding alphabet: full-width-half-maximum (FWHM) of the amplitude profile of each bin, and inter-bin separation. In the case of the HG alphabet, the reported FWHM corresponds to the HG0 (Gaussian) mode, and the inter-bin separation is not defined.}
    \label{tab:param}
\end{table}

\section{Conclusion}
We demonstrated a high-dimensional mode-sorter at the single-photon level based on an mQPG that can programmatically switch encoding alphabet between pulse modes, time bins and frequency bins. Our experimental results show an average measurement fidelity at the single-photon level of up to $0.958\pm 0.030$ in $d=3$ for three MUBs and $0.859\pm 0.131$ in $d=5$ for five MUBs, overcoming the previous dependence of the mQPG on strong output filtering.
The mQPG-based mode sorter not only facilitates the practical realization of HD-QKD across various encoding alphabets, but also has the potential to function as a Hadamard receiver \cite{guha11, rosati16}, enabling quantum communication protocols that go beyond QKD.
By addressing technical challenges such as spectral bandwidth, synchronization, and spectrographic resolution, the measurement fidelity of the system can be improved to scale the mode-sorter to even higher dimensions. 
The presented results show that the mQPG provides an ideal platform for unlocking the benefits of high-dimensional encodings in quantum information science.

\begin{backmatter}
\bmsection{Funding}
This research was supported by the EU H2020 QuantERA ERA-NET Cofund in Quantum Technologies project QuICHE.

\bmsection{Acknowledgments}
The authors thank Abhinandan Bhattacharjee for helpful discussions.

\bmsection{Disclosures}
The authors declare no conflicts of interest.

\bmsection{Data availability}
Data underlying the results presented in this paper are not publicly available at this time but may be obtained from the authors upon reasonable request.
\end{backmatter}

\bibliography{bibliography}

\end{document}


\maketitle

\section{Dimensional scalability}
Here we investigate the scalability of the FFB mode-sorting method to higher dimensions. In $d$ dimensions, this technique involves sorting $d$ frequency bins and their superpositions into $d$ distinct mQPG channels. In this simulated study, we consider an experimental setup analogous to the one presented in the main text, where a waveshaper ``carves'' the pump spectral bins from a fixed broadband laser pulse. In this system, the maximum width of each bin is limited by the initial spectral bandwidth $\Delta\nu_\mathrm{Pump}$ of the laser pulse and by the chosen dimensionality $d$. 

Since broader frequency bins generally lead to a better performance of the mQPG \cite{donohue18, serino23}, this bandwidth constraint is a major source of error in the mode-sorting process. The operation of the mQPG is ideal when the spectral features of the pump pulse (i.e., the width of the pump bins) are significantly larger than the width of the phase-matching function $\Delta\nu_\mathrm{PM}$. For this reason, we study how the ratio between $\Delta\nu_\mathrm{Pump}$ and $\Delta\nu_\mathrm{PM}$ affects the average error of the mode-sorting process by simulating the mQPG projections in each possible mutually unbiased basis of a $d$-dimensional frequency-bin Hilbert space. We set the separation between the frequency bins to approximately 3 times the full-width-half-maximum of each bin.
In the simulation, we assume ideal experimental conditions: perfect resolution for both the pump waveshaper and the output spectrograph, negligible second-order dispersion in the mQPG waveguide, and a constant relative delay between input and pump pulses.

\begin{figure}[h]
    \centering
    \includegraphics[width=.55\linewidth]{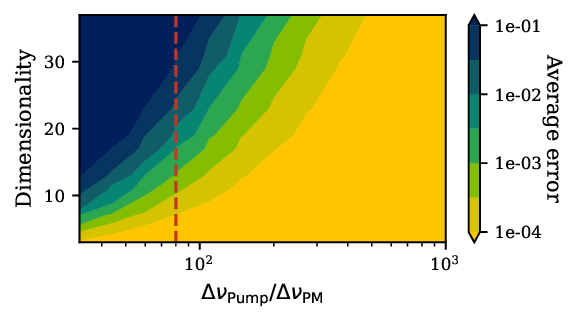}
    \caption{Simulated error of the FFB technique in different dimensions as a function of the ratio between the available pump bandwidth $\Delta\nu_\mathrm{Pump}$ and the phase-matching bandwidth $\Delta\nu_\mathrm{PM}$.}
    \label{fig:dimension-study}
\end{figure}

The simulation results, shown in Fig. \ref{fig:dimension-study}, provide an estimate of the performance the mQPG mode-sorter can ultimately achieve if all technological challenges are addressed. In the top left region of the plot, the mQPG requires too many phase-matching peaks within a narrow spectral range, causing the output channels to overlap and introducing additional cross-talk. The red line indicates the bandwidth ratio of the current experimental setup, which could theoretically operate in 30 dimensions with a measurement error of 10\%. A more broadband laser or a waveguide with narrower phase-matching peaks could enable high-quality mode-sorting in even higher dimensions. 
Overall, our simulation demonstrates that the FFB technique can, in principle, enable high-dimensional operation with extremely low error rates, limited primarily by the technical quality of the experimental components, particularly their resolution and stability.

\section{Spectral representation of the encoding alphabets}
\label{app:spectra}

\begin{figure}[p]
    \centering
    \includegraphics[]{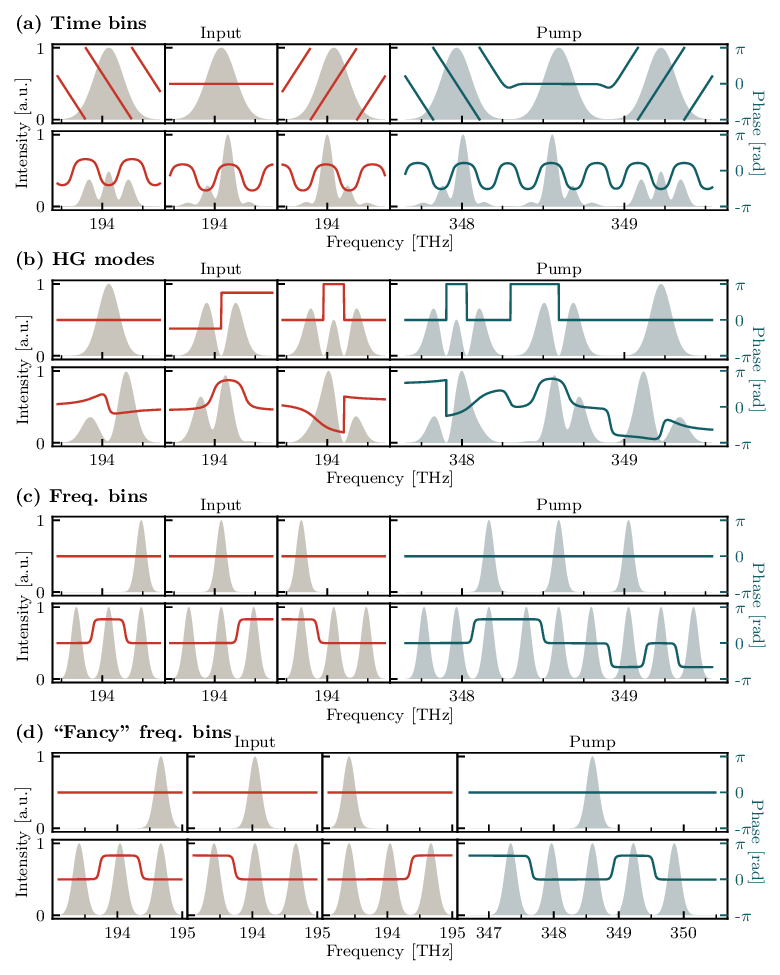}
    \caption{Spectral intensity and phase of the fundamental basis (top) and one of the implemented superposition bases (bottom) of the different encodings used in $d=3$: time bins, HG modes, frequency bins, and FFB approach. The red figures show the different possible input modes, whereas the blue figure on the right shows the pump spectrum that facilitates mode-sorting onto the corresponding basis. 
    }
    \label{fig:spectra3d}
\end{figure}

\begin{figure}[p]
    \centering
    \includegraphics[]{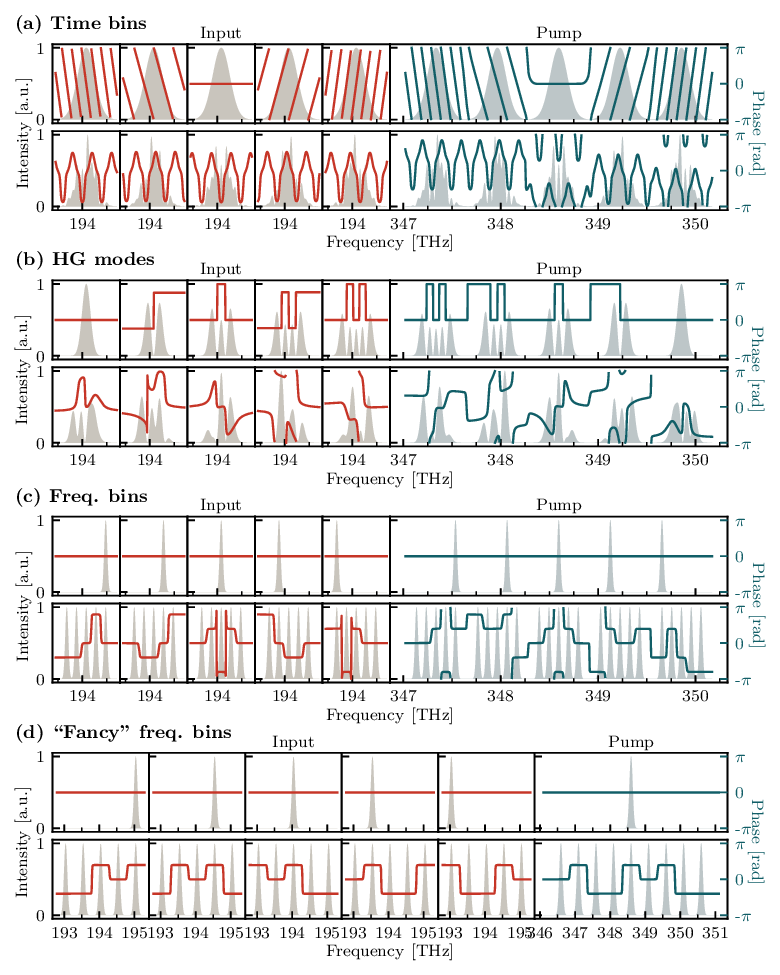}
    \caption{Spectral intensity and phase of the fundamental basis (top) and one of the implemented superposition bases (bottom) of the different encodings used in $d=5$: time bins, HG modes, frequency bins, and FFB approach. The red figures show the different possible input modes, whereas the blue figure on the right shows the pump spectrum that facilitates mode-sorting onto the corresponding basis.}
    \label{fig:spectra5d}
\end{figure}

Figures \ref{fig:spectra3d} and \ref{fig:spectra5d} show, in frequency space, the input and pump modes from the fundamental encoding alphabet and one of the implemented superposition bases in $d=3$ and $d=5$, respectively.

\bibliography{bibliography_supplementary}